\documentclass[aps,prb,twocolumn,superscriptaddress]{revtex4-1}

\usepackage{amsmath, amssymb}
\usepackage{graphicx}
\usepackage{color}

\begin{document}
\title{Nonequilibrium steady-state theory of photodoped Mott insulators}
\author{Jiajun Li}
\author{Martin Eckstein}
\affiliation{Department of Physics, University Erlangen-Nuremberg, 91058 Erlangen, Germany}
\date{\today}

\begin{abstract}
Photodoped states are widely observed in laser-excited Mott insulators, in which charge excitations are quickly created and can exist beyond the duration of the external driving. Despite the fruitful experimental explorations, theoretical studies on the microscopic models face the challenge to simultaneously deal with exponentially separated time scales, especially in multi-band systems, where the long-time behaviors are often well beyond the reach of state-of-the-art numerical tools. Here, we address this difficulty by introducing a steady-state  description of photodoped Mott insulators using an open-system setup, where the photodoped system is stabilized as a non-equilirium steady-state (NESS) by a weak external driving. Taking advantage of the stationarity, we implement and discuss the details of an efficient numerical tool using the steady-state Dynamical Mean-Field Theory (DMFT), combined with the non-crossing approximation (NCA). We demonstrate that these stationary photodoped states exhibit the same properties of their transient counterparts, while being solvable with reasonable computational efforts. Furthermore, they can be parametrized by just few physical quantities, including the effective temperature and the density of charge excitations, which confirms the universal nature of photodoped states indeed independent of the excitation protocols. As a first application, we consider the stationary photodoped states in a two-band Hubbard model with intertwined spin-and-orbital ordering and find a family of hidden phases unknown from the previous studies, implying an apparently unexplored time regime of the relaxation of the intertwined orders.
\end{abstract}

\maketitle

\section{introduction}
The recent years witnessed a surge of interests in the ultrafast dynamics of quantum materials driven by strong laser pulses \cite{basov2017,giannetti2016}. A plethora of experimental studies are carried out on the photoinduced dynamics of transition metal oxides and heterostructures, revealing new possibilities of manipulating material properties on the pico- and even femto-seconds time-scale, such as a putative enhancement of superconductivity \cite{fausti2011, mitrano2016, cavalleri2018}, ultrafast modification of magnetic properties \cite{li2013nature,mikhaylovskiy2015, kirilyuk2010}, and the emergence of photo-induced hidden phases with intertwined spin, orbital, and charge orders \cite{ichikawa2011,stojchevska2014}. 

In general, light-induced dynamics in Mott insulators involves multiple entangled degrees of freedom, and physical processes which occur over orders of magnitude different time scales. One of the most widely established excitation protocol in Mott insulators,  termed photodoping, is the creation of charge excitations, such as doublons (doubly occupied sites) and holons (empty sites) in a single-band system. A laser pulse of femtosecond duration can easily create a significant density of such charge excitations across the Mott gap. These charge excitations can exist well beyond the short duration of the laser pulse,\cite{Iwai2003,Okamoto2010,Mitrano2014} and are decisive for the subsequent non-equilibrium dynamics through their interacting with spin, orbital or lattice degrees of freedom \cite{wall2009, chuang2013, beaud2014, wegkamp2014, mor2017, ligges2018, afanasiev2019, mor2017, dean2016, nembrini2016}. Theoretically, the long lifetime of the charge excitations is explained with a lack of efficient scattering channels to dissipate their large potential energy \cite{sensarma2010,eckstein2011,lenarcic2013,Lenarcic2015}. At the same time, the partial thermalization of doublons/holons inside each Mott-Hubbard bands can be much faster, allowing for the possible formation of quasi-stationary non-thermal states or hidden phases in quantum materials.\cite{werner2012, golez2014, werner2018, li2018nat, peronaci2019,werner2018,golez2017}

Due to this argument, the understanding of photodoped states plays crucial roles in unraveling the complex photo-induced dynamics in realistic systems. For theoretical studies, one common strategy is to explicitly compute the time evolution of the photoexcited model system and examine its physical properties in the long-time limit, where the transient photodoped state becomes quasi-stationary due to the slow charge recombination. Different methods for solving strongly correlated materials, such as Dynamical Mean-Field Theory (DMFT) \cite{aoki2014}, exact diagonalization, and density matrix renormalization group \cite{schollwoeck2004}, have been generalized to the nonequilibrium regime. However, the time scale of the experimentally relevant dynamics can be orders of magnitude longer than the intrinsic time-scales of the electronic systems, such as electron hopping and intraband scattering, which provides a major challenge for microscopic simulations of the real-time dynamics.

Nonetheless, the separation of timescales allows for an alternative method to study the photodoping physics. Since the charge excitations thermalize quickly within the Hubbard bands, and decay on a timescale which is orders of magnitude longer, the experimentally observed photoexcited dynamics can be understood through a quasi-stationary ``non-equilibrium free-energy landscape'' determined by a suitable non-equilibrium control parameter, given by the density of charge excitations in the present case, which gradually evolves as doublons and holons recombine. In practice, this motivates a semiclassical description of the dynamics, such as a Ginzburg-Landau theory. While such an approach is a powerful phenomenological theory and widely used both in theory and for the interpretation of experiments \cite{Nasu2004,beaud2014,stojchevska2014,afanasiev2019,Dolgirev2020,Sun2020}, it is not straightforward to link it to a more quantitative description including the feedback between quantum fluctuations of the electrons and the order parameters, in particular for Mott insulators.  

In this paper, we explore an open-system approach to study the properties of the quasi-stationary photodoped states. Specifically, we apply a weak external driving (through external bath) to compensate the loss of charge excitations due to the slow charge recombination and stabilize the transient photodoped state as a true nonequilibrium steady state (NESS). If the doublon-hole recombination rate is slow, then we can expect two important properties of the resulting NESS: First, we expect that the external driving can be chosen much weaker than the intrinsic energy scales of the system and is still sufficient to maintain a nonzero excitation density. Second, a doublon-hole pair inserted into the system from the bath remains in the system much longer than the intra-band thermalization time, so that the NESS should become universal. It is then largely independent of the detailed properties of the  bath, and dependent on only few effective parameters, which can be taken as the control parameters in a nonequilibrium phase diagram. We implement this protocol with a coupling to carefully chosen fermion reservoirs, which steadily inject doublons and holons into a Mott insulator, without breaking the symmetries of the model.  We will then demonstrate that this bath-doping protocol indeed produces steady-states with quantitatively the same properties as the laser excited systems, and discuss a first application to the spin-orbital-ordered two-band Hubbard model. 

Another way to explain the approach is to note that deep in the Mott phase the double occupancy becomes an almost conserved quantity, since the upper and lower Hubbard bands are energetically well separated. The approach to stabilize a universal NESS is then similar in spirit to the activation of almost conserved quantities in near integrable systems \cite{lange2017}. The NESS formalism should be also contrasted with the idea used successfully in diagrammatic weak-coupling calculations for insulators to impose certain nonequilibrium distribution functions in the conduction and valance band (see, e.g., Ref.~\onlinecite{wegkamp2014} for an example within the  $GW$ formalism), or assume a Fermi distribution with separate chemical potentials for the electrons in the conduction and valance band. In the bath doping, both non-thermal distribution functions and the modification of the spectrum due to the modified distribution and correlations are determined self-consistently, which make it suitable for the application to strongly correlated systems.

The article is organized as follows: In Sec.~\ref{model}, we discuss the Hubbard model and the fermion bath-coupling. We discuss the details of a specific bath setup to create the non-equilibrium steady-state (NESS) containing excess charge excitations. We also elaborate on the solution of the model using nonequilibrium DMFT and the steady-state NCA impurity solver. In Sec.~\ref{pump}, we show that the bath-coupling pumps up charge excitations and discuss the universality of these states independent of the bath details. Sec.~\ref{cmp} then concentrates on a systematic comparison between the stationary and the transient photodoped states, created by bath-coupling in the present setup and from real-time DMFT simulation, respectively. Sec.~\ref{2band} applies the bath-doping protocol to a two-band Hubbard model with intertwined spin and orbital orders and discusses the relation between the bath-doped states and the transient photodoped states. Sec.~\ref{conclusion} is conclusion and outlook.

\section{model and method}
\label{model}

As the minimal setup to illustrate the steady-state formalism, we consider a one-band Hubbard model defined on the Bethe lattice of infinite coordination number, described by the following Hamiltonian 
\begin{align}
H=-t_0\sum_{\langle ij\rangle\sigma}c^\dag_{i\sigma} c_{j\sigma} + U\sum_i n_{i\uparrow}n_{j\downarrow}+\mu\sum_i n_i,
\label{ham}
\end{align}
where $c_i$ annihilates the lattice electron at site $i$ and $\langle ij\rangle$ runs over all pairs of neighbors. The model features a semielliptic one-particle density of states $\rho(\omega)=\sqrt{4t_0^2-\omega^2}/\pi t_0^2$ (bandwidth $4t_0$) and is a minimal model describing the Mott physics \cite{georges1996}. For large interaction $U/t_0$ and at half-filling, the low temperature phase is antiferromagnetically ordered \cite{georges1996}. In our calculations the half-filling condition is imposed by the condition that the chemical potential $\mu=-U/2$. We use $U=8.0t_0$ unless otherwise stated; $\hbar=1$ is set throughout the paper.

\begin{figure}
\includegraphics[scale=1.3]{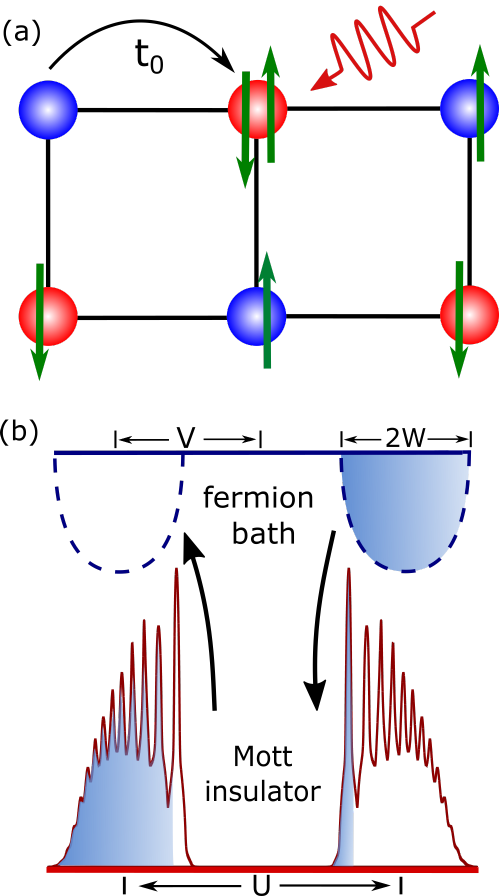}
\caption{
Illustration of real laser-induced photo-doping and the auxiliary bath doping protocol: 
(a) In a Mott insulator, photo-excitation creates charge excitations. This effectively dopes both doublons and holons into the system. (b) Coupling a Mott insulator to 
suitably chosen
fermion baths can lead to 
non-equilibrium doping effects similar to the photo-excitation. The spectral function is calculated for the Hubbard model with $U=8, \beta=12.5$ in the
antiferromagnetic ground state and averaged over spins. The blue shades schematically indicate the charge occupation. }
\label{fig1}
\end{figure}

As discussed in the introduction, we intend to maintain a stationary photodoped state with fermion-bath coupling. Specifically, we weakly couple the system to two fermion baths which are individually in equilibrium, and have otherwise identical density of states except for different chemical potentials and energy shifts. To be concrete, the spectra of the baths are shifted by $\pm V$, with the intention of matching their bands with the upper or the lower Mott-Hubbard bands of the lattice system, as schematically depicted in Fig.~\ref{fig1}. The Hamiltonian of fermion baths reads,
\begin{align}
H_{\rm bath} &= g\sum_{is\alpha} (d^\dag_{is\alpha}c_i + {\rm h.c.})+\nonumber\\
&\quad+\sum_{is\alpha}(\epsilon_\alpha+V_s)d^\dag_{is\alpha}d_{is\alpha},
\end{align}
where $g$ is coupling constant and $s=U$ or $L$ corresponds to upper and lower bath. 

Due to the particle-hole symmetry of the hamiltonian~\eqref{ham}, we assume a symmetric bath setup, that is, for upper ($s=U$) and lower ($s=L$) bath we impose $V_U= V$ and $V_L=-V$, respectively. For simplicity, we assume the upper bath is full and the lower bath is empty. In the resulting NESS, the upper bath should mainly inject electrons into the upper Hubbard band, forming doublons, 
i.e., doubly occupied lattice sites. The lower bath, on the other hand, absorbs electrons out of the lower Hubbard band, forming holons, i.e., empty lattice sites. Charge excitations are constantly created in the lattice, cancelling the spontaneous recombination of 
doublons and holons. The excess energy is also dissipated through the baths, thus producing a stationary doublon and hole-doped state, in analogy to a photodoped state.

In the following, we will generally consider two types of spectra for the fermion baths, namely the semi-elliptic and the square spectra. After integrating out the bath degrees of freedom, the semi-elliptic bath yields a local self-energy with $-{\rm Im}\Sigma^r_{\rm bath}(\omega)=\pi g^2 D_{\rm bath}(\omega)$, where $D_{\rm bath}(\omega)=\sum_{\alpha,s}\delta(\omega-\epsilon_\alpha)$ is the local density of states of the bath. The semi-elliptic bath yields a self-energy $-{\rm Im}\Sigma^r_{\rm bath}(\omega)=\Gamma\sum_s \sqrt{1-(\omega-V_s)^2/W^2}$ while the square bath gives rise to $-{\rm Im}\Sigma^r_{\rm bath}(\omega)=\Gamma/2\sum_s\theta(\omega-V_s-W)\theta(W-\omega+V_s) $ with Heaviside step function $\theta$. Here the damping constant is defined as $\Gamma=g^2/W$ for both cases.

We use nonequilibrium Dynamical Mean-Field Theory (DMFT) to solve the model. In general, DMFT maps the lattice problem to a single-impurity Anderson model (SIAM) with a self-consistently determined bath and is exact in the infinite dimensional limit \cite{georges1996}. In particular, it is exact for model \eqref{ham} on the Bethe lattice of infinite coordination number. The steady-state nonequilibrium DMFT is recently developed to study systems under constant or periodic driving \cite{aoki2014,joura2008, li2015,matthies2018, murakami2018}. In the steady state, due to the time-translational invariance of the problem, all Green's function $G^{r,<}(t,t')=G^{r,<}(t-t')$ can be Fourier transformed to the frequency domain. One can firstly integrate out the bath degrees of freedom and map the lattice problem to the following SIAM, 
\begin{align}
&S_{\rm imp}=
-i\sum_{\sigma\sigma'}\int_\mathcal{C} dtdt' c_{\sigma}^\dag(t)\Delta_{\sigma\sigma'}(t-t')c_{\sigma'}(t')\nonumber\\
&-i\int_\mathcal{C} dt\left[\sum_\sigma c_{\sigma}^\dag(t)(-i\partial_t-\mu)c_\sigma(t)+ Un_{\uparrow}(t)n_{\downarrow}(t)\right],
\label{imp}
\end{align}
where $\mathcal{C}$ denotes the Keldysh contour and the hybridization function $\Delta_{\sigma\sigma'}(t-t')=\Delta^{latt}_{\sigma\sigma'}(t-t')+\Sigma_{{\rm bath}}(t-t')\delta_{\sigma\sigma'}$ encodes both the self-consistent bath $\Delta^{latt}$ and a contribution 
$\Sigma_{{\rm bath}}$ from the real fermion baths. 
For the Bethe lattice, the self-consistent bath is given by $\Delta^{latt}_{\sigma\sigma'}(t-t') = t_0^2G_{{\rm loc},{\sigma\sigma'}}(t-t')$, where $G_{\rm loc}$ represents the Green's function of the ``central" site of the Bethe lattice. The real bath contribution is obtained after integrating out the bath degrees of freedom, as noted above. If a bath is empty or full, one simply has to assumes the relation between the lesser or greater components  and the retarded (advanced)  component 
\begin{align}
&\Sigma_{{\rm bath}}^>=0,\,\,\, \Sigma_{{\rm bath}}^<=\Sigma_{{\rm bath}}^{a}-\Sigma_{{\rm bath}}^{r} \text{~(empty bath)},
\\
&\Sigma_{{\rm bath}}^<=0,\,\,\, \Sigma_{{\rm bath}}^>=\Sigma_{{\rm bath}}^{r}-\Sigma_{{\rm bath}}^{a} \text{~(filled bath)}.
\end{align}
(More generally, one can also use partially filled bath with a separate chemical potential. In the present paper, this is used only for the extremely strongly doped case, discussed in the appendix.) In this work, the auxiliary SIAM is solved by a self-consistent strong coupling expansion up to the lowest order, a.k.a. non-crossing approximation (NCA) \cite{coleman1984,eckstein2010}. The real-time implementation of NCA can be readily generalized to the steady-state \cite{murakami2018prl, murakami2018}. The implementation of the NCA is discussed with more details in the appendix. In general, one starts with an arbitrary initial guess, and self-consistently update the solution until convergence, i.e., when $G_{\rm imp}=G_{\rm loc}$.

In the following, we will examine the spectral function $A(\omega)=-\operatorname{Im} G^r_{\rm loc}(\omega)/\pi$, which indicates the local density of states, and the distribution function $f(\omega)=-\operatorname{Im} G^<_{\rm loc}(\omega)/2\operatorname{Im} G^r_{\rm loc}(\omega)$, which represents the local occupational probability for electrons and is the Fermi-Dirac distribution in equilibrium. Equivalently, the distribution can be analyzed via the occupied density of states, $-iG^<_{\rm loc}(\omega)=2\pi A(\omega)f(\omega)$. 

\section{Pumping charge excitations in a Mott insulator}
\label{pump}
In this section, we show the fermion bath-coupling effectively pumps up charge excitations in an insulating system. This can be demonstrated by analyzing the non-thermal distribution function $f(\omega)$ in the bath-coupled state. As discussed above, the charge excitations, once created, should decay slowly in terms of electron hopping time ($t_0^{-1}$), so that a very weak coupling to the fermion baths should suffice to maintain a stationary photodoped state. In such a case, one may expect the resulting stationary photodoped state exhibits universal properties independent of the details of the bath, such as its density of states and occupation. We will confirm this universality in the following.

\subsection{The universality in the antiferromagnetic phase}
We start with the photodoping in the antiferromagnetic ground state of a half-filled Hubbard model. To be specific, we couple the system with a full upper fermion bath as well as an empty lower fermion bath, both of square-shaped density of states. The calculation is started with an \emph{equilibrium} initial guess of inverse temperature $\beta=12.5$, and the $V_{U/L}$ are chosen so that the overlaps between the Hubbard bands and the fermion baths are small, and the photodoping level is moderate. Note that, for extremely strong photodoping, the solution can in general depend on the initial guess \footnote{see appendix for more details}. 

The effect of the bath-coupling is best demonstrated by looking at the distribution function $f(\omega)$. In an equilibrium fermion system, fluctuation-dissipation theorem imposes $f(\omega)$ to be a Fermi-Dirac distribution. However, $f(\omega)$ can exhibit enhanced occupation near the upper Hubbard band and reduced occupation near the lower band upon photodoping, indicating the distribution of doublons and holons, respectively. 

\begin{figure}
\includegraphics[scale=0.7]{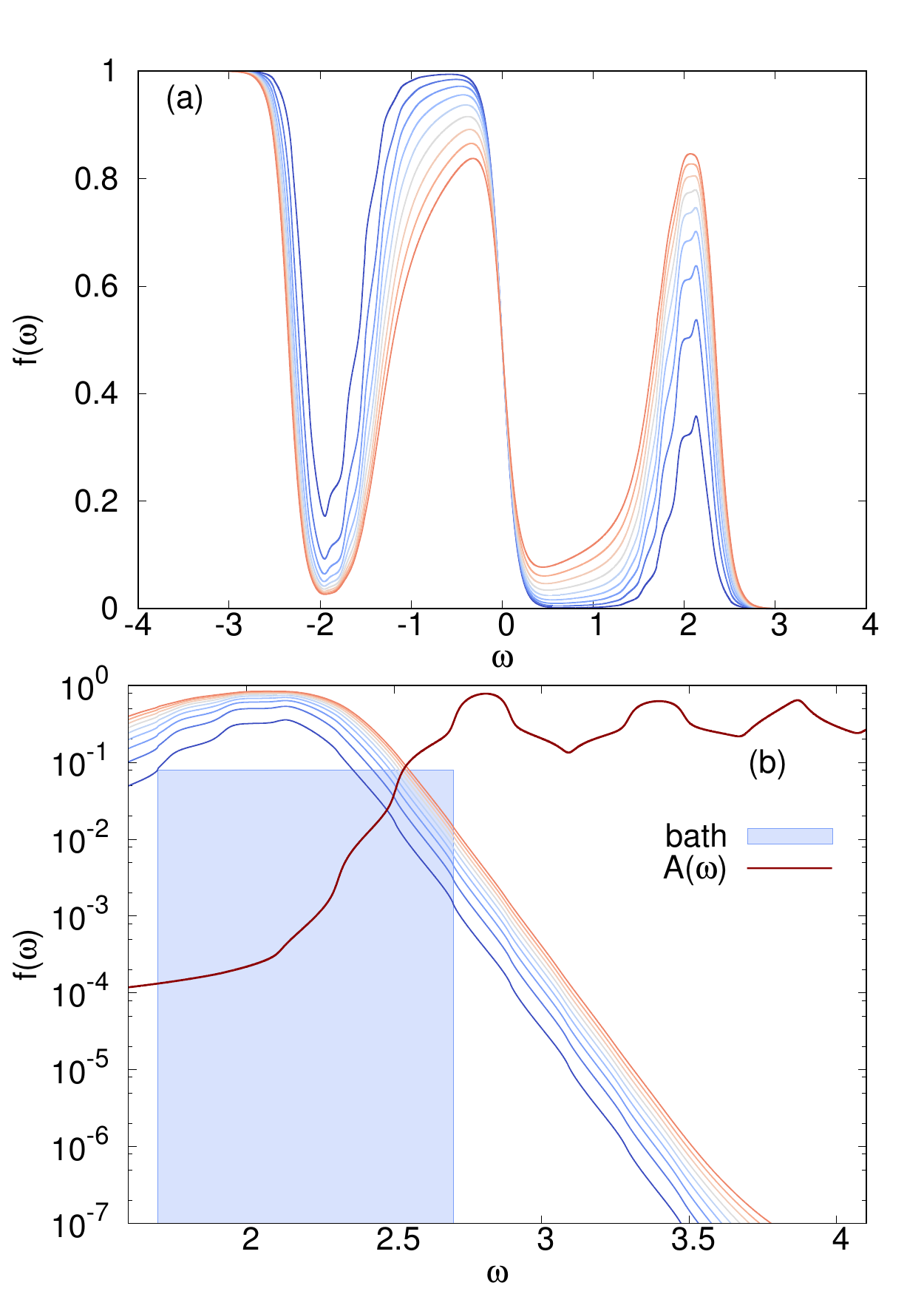}
\caption{
Distribution function $f(\omega)$ of bath-doped systems in AFM phase. The minority spin is shown. The dark-red line is the spectral function  $A(\omega)$. The blue-shaded region indicates the fermion-bath occupation. The bath coupling is changed in the range $\Gamma=0.2,0.4,0.6\ldots, 1.8 \times10^{-4}$, shown as solid lines from blue to red colors, respectively.}
\label{fig2}
\end{figure}

As shown in Fig.~\ref{fig2}, the distribution function $f(\omega)$ evolves continuously out of equilibrium for increasing bath-coupling, and a peak (dip) gradually emerges at the bottom (top) of the upper (lower) Hubbard band. Moreover, a closer look at the tail of the doublon peak exhibits evidently an exponential decay, as shown in panel (b) of the figure. Therefore, a well-defined effective temperature $T_{\rm eff}=1/\beta_{\rm eff}$ of charge excitations can be determined from the slope $\partial_\omega \log(f(\omega))\sim -\beta_{\rm eff}$ above  the effective ``Fermi level" of the charge excitations (doublons in the figure). Analogous behaviors are observed for holes in the lower Hubbard band. Interestingly, the effective temperature only changes slightly as $\Gamma$ increases. On the other hand, the effective temperature $T_{\rm eff}$ can be significantly modified by choosing different $V$ of baths, since the latter affects the energy flow into/out of the system. 

Note that the rapidly varying form of the distribution function in the frequency range where the spectrum $A(\omega)$ vanishes has no consequence for physical observables. Therefore, the data show that the distribution function $f(\omega)$ in the relevant frequency range $\omega\gtrsim 2.2$ with $A(\omega)>0$ (see the dark-red curve in Fig.~\ref{fig2}b) is essentially described by only two parameters, i.e., the total occupation in the upper band, and the effective temperature. In particular, $f(\omega)$ has no detailed resemblance to the sharp edge in the density of states. Below we will see that the same holds for other bath density of states. Such a universal distribution of doublons (holons) is consistent with the fact that, in the AFM phase, charge excitations couple strongly with the long-range-ordered spin moments, and the AFM order acts as a reservoir to assist the relaxation of the excitations. (In equilibrium, this charge-spin coupling results in the sub-structure in the spectral function in Fig.~\ref{fig2}(a).) Once a doublon-hole pair is inserted from the bath, it therefore thermalizes much faster than the recombination time, and detailed memory on the bath density of states is lost in the steady state.

\subsection{The non-universality in the paramagnetic phase}
\begin{figure}
\includegraphics[scale=0.7]{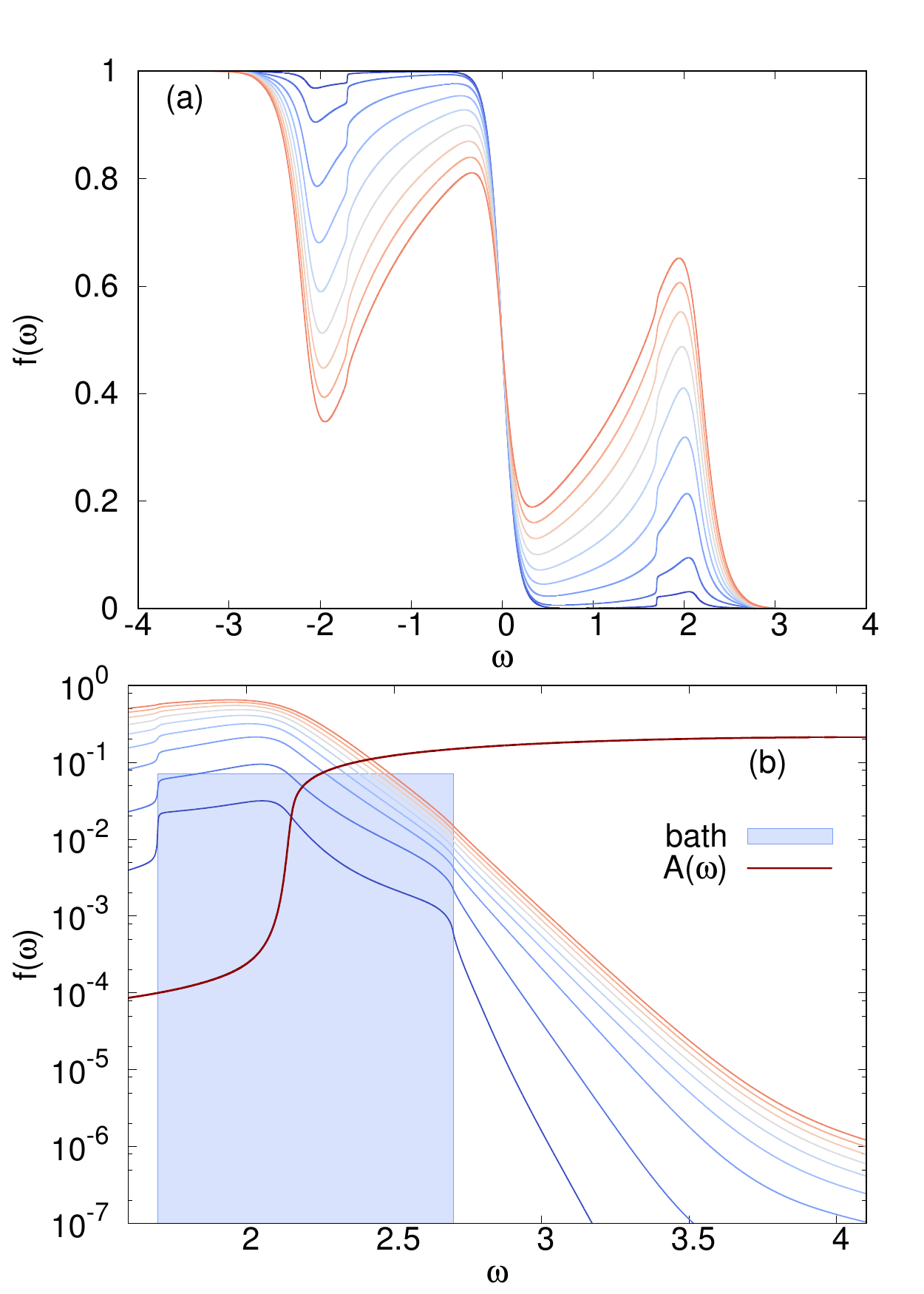}
\caption{
Distribution function of bath-doped systems in PM phase. The dark-red line is the spectral function $A(\omega)$. The chosen $\Gamma$'s and the color scheme are identical to Fig.~\ref{fig2}. }
\label{fig3}
\end{figure}

The situation changes in the paramagnetic phase. As shown in Fig.~\ref{fig3}, the distribution function is strongly affected by the square-shaped spectrum of the fermion baths at weak bath coupling. The distribution function even forms shoulders at the edges of the bath spectrum, which can be observed in both panels of Fig.~\ref{fig3}. These shoulders shift positions following the fermion baths if $V$ is changed. For stronger bath-doping, e.g. $\Gamma\gtrsim 1.0\times10^{-4}$, the non-universal features are suppressed due to the higher density of charge excitations, which results in enhanced quasiparticle scattering and intraband thermalization. We note that even in this regime the bath coupling is orders of magnitude smaller than other major energy scales in the system. We also observe that the charge excitations in PM phase are generally hotter than those in the AFM phase, indicated by larger effective temperatures ($T_{\rm eff}$). This may be explained by the absence of long-range order, which leads to less efficient relaxation of charge excitations than in the AFM phase. 

The lack of universality is again consistent with the study of ``intra-band'' thermalization of excitations in the paramagnetic phase pf the Hubbard model. In previous time-dependent studies, it has been observed that the relaxation of charge excitations in the paramagnetic Mott phase is almost stuck \cite{eckstein2014, werner2014}, and retains a detailed memory on the spectrum of the pump laser pulse. In part, this is understood as an artifact of the local approximation within DMFT, because this interaction of electrons wth long range charge-fluctuations \cite{golez2015} or spin fluctuations \cite{eckstein2016,bittner2020} would introduce a fast relaxation scale. One could therefore expect that the proper incorporation of such interactions beyond DMFT would yield a universal bath-doped state also in the paramagnetic phase.

In summary, the distribution function of charge excitation generally exhibits universal features under different bath-coupling, suggesting a well-defined inverse effective temperature $\beta_{\rm eff}$ measured by the exponential tails in $f(\omega)$, or more generally for stronger doping a distribution of the form $f(\omega)=(\exp(\beta_{\rm eff}(\omega-\mu_{\rm eff}))+1)^{-1}$. This leads to a well-defined stationary photodoped state, which can potentially be parametrized by a few physical quantities including $T_{\rm eff}$. 
 
\section{The nonequilibrium phase diagram}
\label{cmp}

\subsection{Comparison to time-evolved states}

\begin{figure}
\includegraphics[scale=0.9]{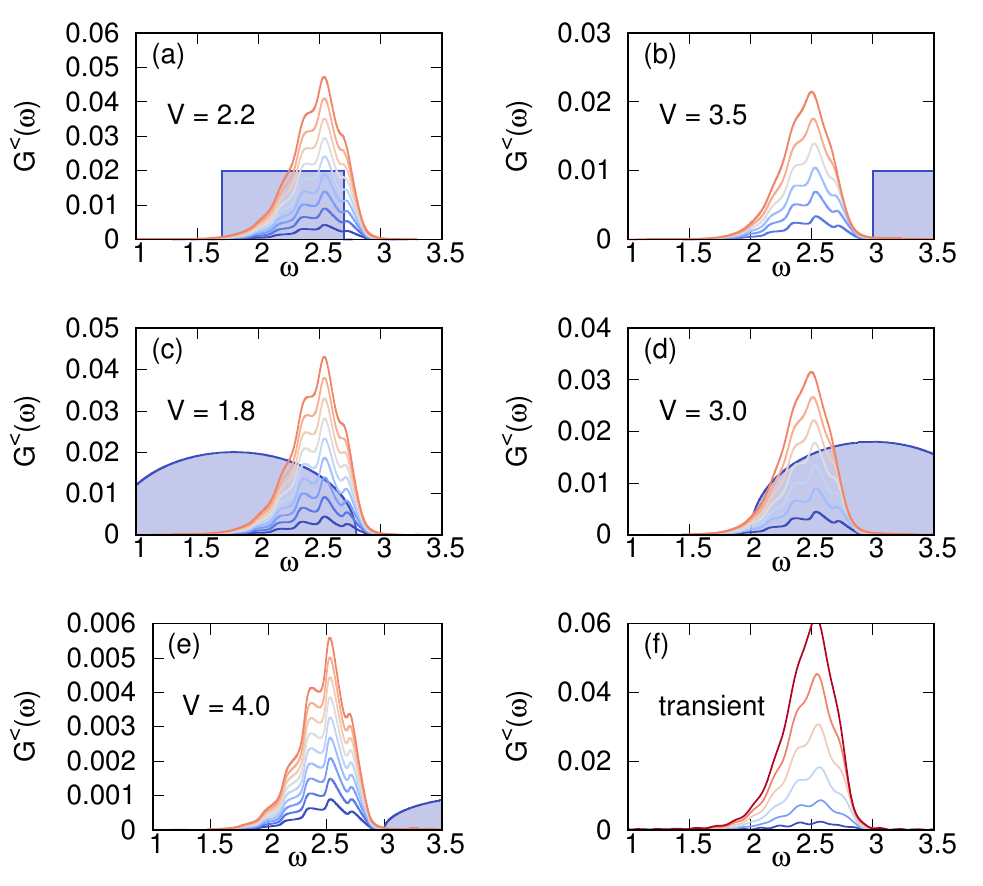}
\caption{Charge excitations induced in the one-band Hubbard model through fermion-bath coupling. Lines with colors from blue to red show occupied density of states $-iG^<(\omega)$ in the upper Hubbard band (for minority spin). The blue areas indicate the shape and position of the baths. Different panels show the distribution for baths with different shapes (square or semi-elliptic) and offsets $V$. The colors from blue to red correspond to increasing bath copulings. The dark-red line $A(\omega)$ indicates the spectrum function of the upper Hubbard band of the minority spin. The panel of transient states shows the distribution created by an electric pulse in the real-time DMFT simulation. The $E_0$ is varied from 0.1, 0.2, \ldots, 0.6. For the steady-states, the damping $\Gamma=$ (a) $0.2,0.4,\ldots,1.8 \times 10^{-4}$, (b) $0.2,0.4,\ldots,1.2 \times 10^{-5}$, (c) $0.2,0.4,\ldots,1.8 \times 10^{-4}$, (d) $0.2,0.4,\ldots,1.4 \times 10^{-5}$, (e) $0.4,0.6,\ldots,2 \times 10^{-6}$.}
\label{fig4}
\end{figure}

To establish the physical relevance of the stationary photodoped states, we compare their properties with a transient photodoped system excited by ultrafast pulses. We first confirm that the distribution of the charge excitations in these stationary states are similar to  that of the transient states. To be concrete, we consider the antiferromagnetic phase (AFM) at inverse temperature $\beta=12.5$ and $U=8.0$. When baths are attached to the system, we find charge excitations accumulate at the bottom (top) of the Upper (Lower) Hubbard bands. Fig.~\ref{fig4} shows the occupied density of states $-iG^<_{\rm loc}(\omega)=2\pi A(\omega)f(\omega)$ at the bottom of the upper Hubbard band for the minority spin, verifying again that the distribution is insensitive to the details of the fermion baths. Similar behaviors are observed for holons in the lower band. In this figure, the bath-coupling $\Gamma$ is varied up to the nonequilibrium phase transition to a paramagnetic phase and is generically of the order of magnitude $\lesssim 10^{-4}$, being much smaller than other energy scales in the system.

These steady-state results are compared with the real-time simulations on the same model \eqref{ham} using nonequilibrium DMFT. Specifically, the equilibrium ground state of $\beta_{\rm eq}=12.5$ is disturbed by a short electric field pulse $E(t)=E_0\sin(\Omega t)\theta(t)\theta(T-t)$ where $T=5.0$, $\Omega=2\pi$, and $\theta(t)$ is the Heaviside step function. The amplitude $E_0$ is varied to reach different photodoping levels. Following the pulse, the system evolves into a quasi-steady state after about $\sim40$ $t_0^{-1}$, and the occupied density of states $G^<(t,\omega)=\int ds\,e^{i\omega s}G^<(t+s/2, t-s/2)$ is shown for time $t=40.0$ for different amplitudes $E_0$, see Fig.~\ref{fig4}(f). 

The distribution for the time-evolved state apparently bears a resemblance to the stationary cases created by bath-coupling. To make this statement quantitative, and show that the properties of the bath-doped and photo-doped state are the same, we now aim to identify the suitable control parameters to scan a phase diagram of the non-equilibrium steady states, and show that the properties of the time-evolved state are in fact reproduced  by a point in this phase diagram.

\subsection{Antiferromagnetic order parameter}

\begin{figure}
\includegraphics[scale=0.67]{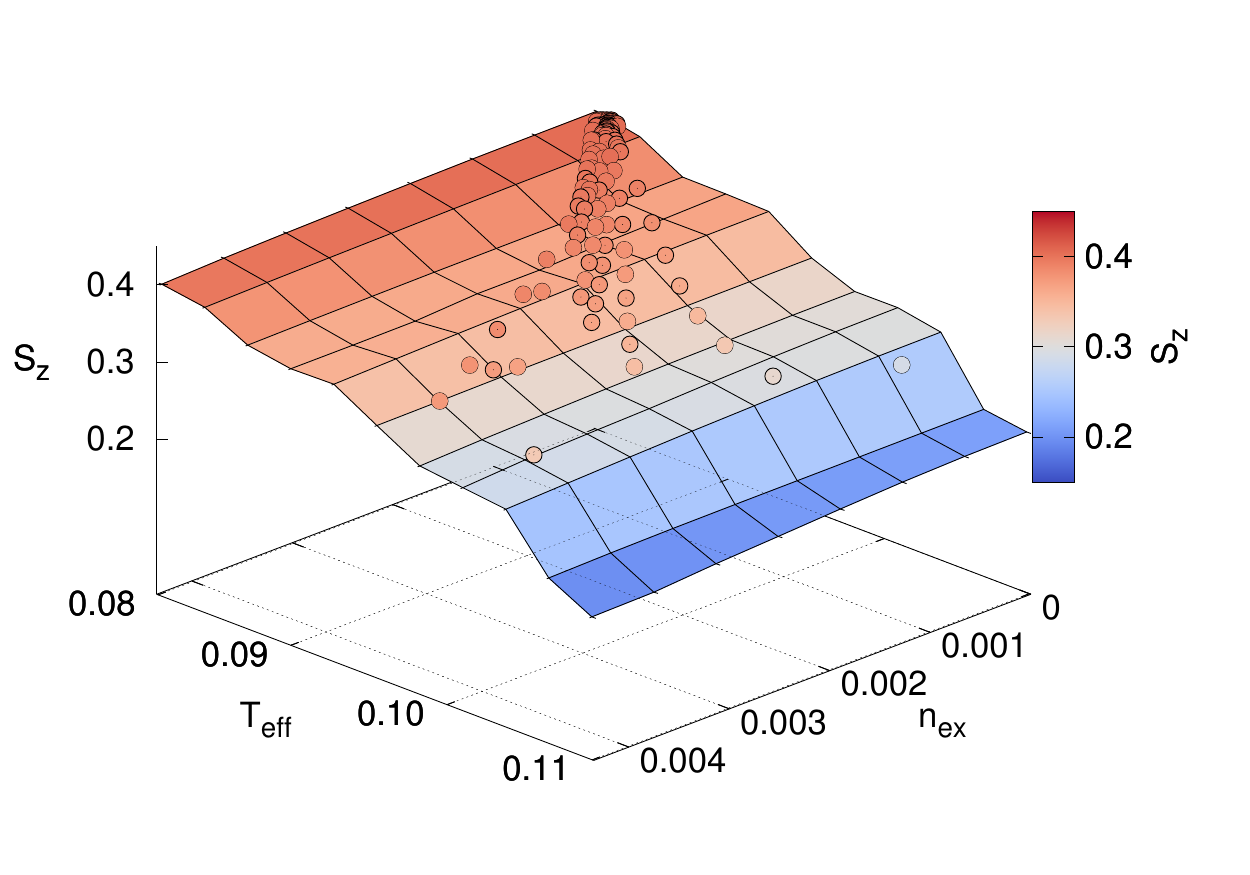}
\caption{The antiferromagnetic order parameter $S_z(T_{\rm eff},n_{\rm ex})$ in equilibrium and nonequilibrium. In this plot, we have chosen $U=8.0$ and $\beta=12.5$ for the equilibrium reference system in an antiferromagnetic phase. The points show $S_z$ in the nonequilibrium steady state with $\Gamma=0.5,1.0,\ldots,5.0\times10^{-7}$, plotted against $n_{ex}$ and $T_{\rm eff}$ as defined in the text. The surface shows equilibrium $S_z$ for different temperatures and chemical potentials. In the equilibrium case, we define $n_{\rm ex} = \frac{1}{2}|n_\uparrow+n_\downarrow-1|$. The factor $1/2$ is introduced to fairly compare with the bath-doped system where both doublon and holons are present.} 
\label{fig5}
\end{figure}

In the AFM phase, the presence of charge excitations can significantly reduce the ordered moment (spin polarization) $S_z$. To quantify this effect in the stationary photodoped states, we first define the excitation density $n_{\rm ex}$ as the increased value of the double occupancy due to bath-coupling, i.e., $n_{\rm ex}=d(\Gamma)-d(0)$, with the double occupancy $d=\langle n_{\uparrow}n_{\downarrow}\rangle$. Furthermore, one can obtain the effective temperature $T_{\rm eff}$ by fitting the exponential tail of the distribution function $f(\omega)$, as discussed in the previous sections. Since the stationary photodoped states exhibit universal features for the charge distribution, one can speculate the existence of the function $S_z(T_{\rm eff}, n_{\rm ex})$, which maps the parameter tuple $(T_{\rm eff}, n_{\rm ex})$ to the AFM spin order in the photodoped state \cite{werner2012}. For concreteness, we stick to semi-elliptic fermion baths and systematically change $V$ and $\Gamma$ to sample this function, see the symbols Fig.~\ref{fig5}. Although the evaluation of $T_{\rm eff}$ rather sensitively depends on numerical errors, the figure clearly implies the possible existence of a single-valued $S_z(T_{\rm eff}, n_{\rm ex})$, suggesting the stationary photodoped states can be parametrized by $n_{\rm ex}$ and $T_{\rm eff}$. Because a similarity between photodoped and chemically doped (equilibrium) Mott insulators has been discussed in previous works \cite{werner2012}, it is also worthwhile to show in Fig.~\ref{fig5} the corresponding equilibrium function $S_z^{\rm eq}(T, n_{\rm ex})$
which is controlled by temperature and chemical potential, and is plotted as the surface in Fig.~\ref{fig5} for comparison. Also here we note a similarity, that will be analyzed below.

To quantitatively confirm the existence of a ``single-valued'' manifold of states, we note that, in addition to the reduction of $S_z$, the presence of charge excitations also significantly changes the single-particle spectral function \cite{sangiovanni2006}. Indeed, the hopping of doublons and holons in an AFM background leads to trails of defects, resulting in energy transfer into the ordered local-spin moments \cite{lenarcic2013,balzer2015}. This process has two consequences: (i) the AFM order dynamically obtains energy from the charge excitations, leading to increase of the ``spin temperature" and a (partial) melting of the order; (ii) doublons and holes experience an effective potential proportional to their hopping distance, giving rise to so-called spin-polaron peaks in the spectral function \cite{sangiovanni2006}. These peaks already appeared in Fig.~\ref{fig1}(b). 

\begin{figure}
\includegraphics[scale=0.7]{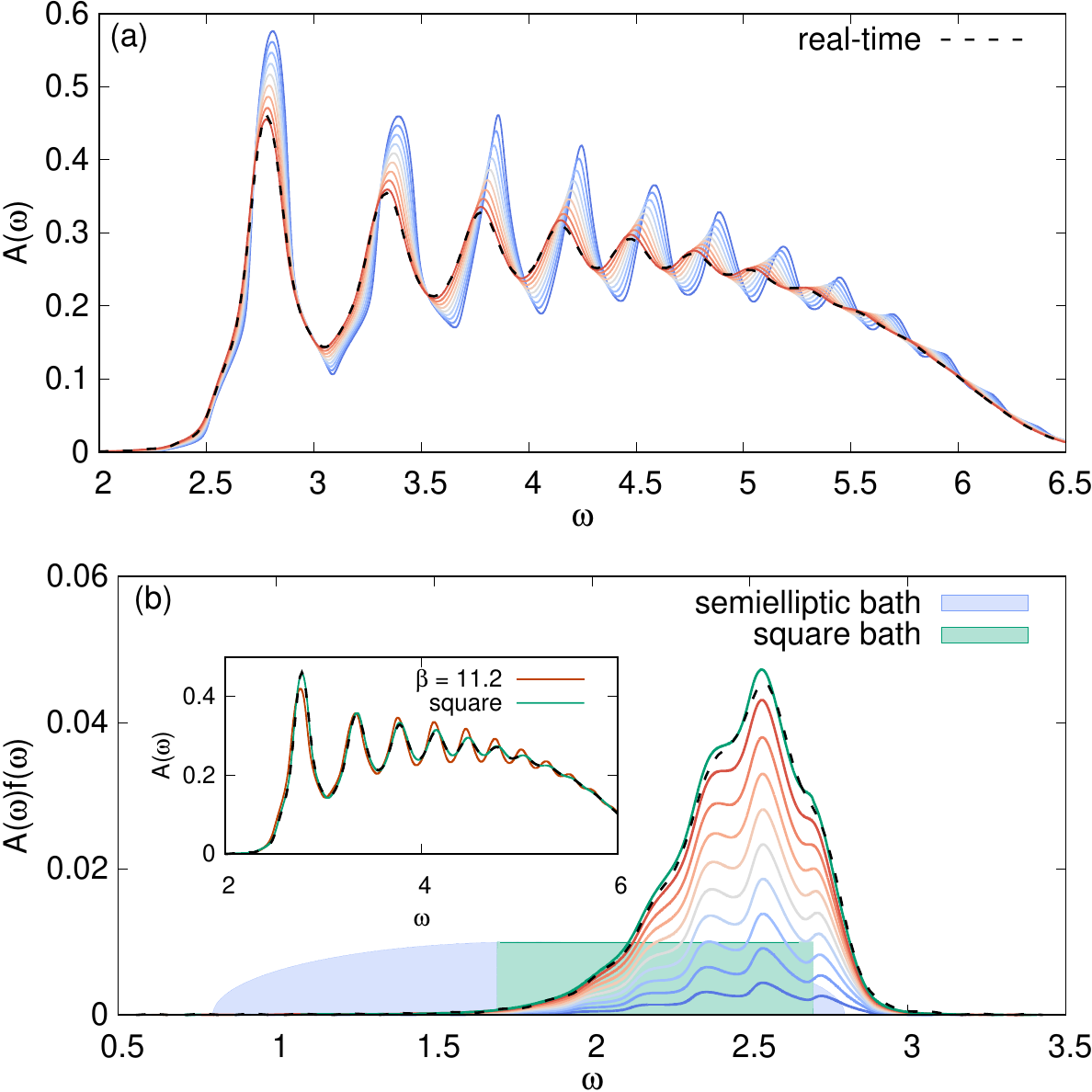}
\caption{(a) Spectral function $A(\omega)$ and (b) occupied spectrum $A(\omega)f(\omega)$ under various bath coupling $\Gamma/10^{-5}=0.5,1.0,\ldots,4.5$ from blue to red. The dashed line is the result of real-time simulation. It roughly fits the red curve with $\Gamma=4.5\times10^{-5}$. The inset of (b) compares the real-time solution, the steady-state with square bath coupling, and the equilibrium state of $\beta=11.2$ at half-filling, which all have rather close AFM order parameters. $W=1.0$ and $V=1.8$ is used for the bath-coupled steady-state system.}
\label{fig6}
\end{figure}

This allows for a unique opportunity to quantitatively compare the stationary and transient photodoped states. Indeed, one should expect that a similar parametrization of $(T_{\rm eff}, n_{\rm ex})$ should exist for the transient photodoped states, and for the same parameters, the spectral function $A(\omega)$ and the occupied density of states $-iG^<(\omega)=2\pi A(\omega)f(\omega)$ should also be the same for the stationary and transient photodoped states if the steady-state theory is valid. For this reason, a series of stationary photodoped states are obtained by varying $\Gamma$, finally reaching $n_{\rm ex}=0.0148$ and $S_z=0.370$, with spectral functions plotted in Fig.~\ref{fig6}. The spectral functions are compared against a transient state of $n_{\rm ex}=0.0149,S_z=0.367$ in the long time limit, which is excited by the electric pulse described above.
Upon increasing $\Gamma$, the spin polaron peaks for the stationary states damp out more and more strongly and, at $\Gamma=4.5\times10^{-5}$, they eventually become identical to the transient state (black dashed line). Moreover, the same result is obtained when square baths are used to excite the system, as indicated by the Green curves in Fig.~\ref{fig6}(b). With the same $n_{\rm ex}$ and $S_z$, the system always shows essentially identical spectral function and distribution of charge excitations, which confirms the single-valuedness of the mapping $S_z(T_{\rm eff}, n_{\rm ex})$, as the distribution in Fig.~\ref{fig6}(b) determines the $T_{\rm eff}$. We stress that the observation is physically reasonable because a bath-coupling $\Gamma$ down to $10^{-5}t_0$ should not affect most of the fast electronic processes in the photodoped system, such as doublon(holon) scattering and the charge-spin interaction. 

In addition, we compare both the transient and stationary photodoped states with an equilibrium state \emph{at half-filling} with a similar spin order $S_z=0.368$. This state is reached by increasing temperature to $\beta=11.7$. As shown in the inset of Fig.~\ref{fig6}(b), the equilibrium spectrum is distinct from the non-equilibrium cases. This indicates that the temperature effect alone cannot explain the spectral features of a photodoped system, and the presence of nonthermal charge excitations is crucial to describe the photodoping physics. This has already been noted in Ref.~\onlinecite{werner2012}, by comparing chemically doped to photo-doped states. Finally, it is worth noting that, although we showed the stationary photodoped states are reasonable approximations of the corresponding transient states, it remains an open question whether the manifold of stationary states cover all the physical scenarios under appropriate bath parameters.

\section{Photo-doping in a two-band Mott insulator}
\label{2band}
So far we have concentrated on the photodoped one-band Mott insulator, while multiple bands are often relevant to the ultrafast dynamics in experimental systems. Indeed, the steady-state theory carried out here provides a promising tool to access the long-time behavior of these multi-band systems, due to a significant reduction of computational costs. In this section, we consider one paradigmatic example, the photodoping in a three-quarter filled ($n=3$) two-band Hubbard model with $e_g$ orbital degeneracy and cubic lattice symmetry. This model is widely used to study the intertwined orders in transition metal compounds, in particular the spin and orbital ordering in KCuF\textsubscript{3}, and is shown to form a hidden phase under photodoping \cite{li2018nat}. The model can be summarized as below,
\begin{align}
H&=U\sum_{i\ell}n_{i\ell\uparrow}n_{i\ell\downarrow}+\sum_{i,\sigma\sigma',\ell\neq\ell'}(U'-J_H\delta_{\sigma\sigma'})n_{i\ell\sigma}n_{i\ell'\sigma'}\nonumber\\
&+J_H\sum_{i,\ell\ne\ell'}(c^\dag_{i\ell\uparrow}c^\dag_{i\ell\downarrow}c_{i\ell'\downarrow}c_{i\ell'\uparrow}+c^\dag_{i\ell\uparrow}c^\dag_{i\ell'\downarrow}c_{i\ell\downarrow}c_{i\ell'\uparrow})\nonumber\\
&-t_0\sum_{\langle ij\rangle\ell\ell'\sigma}{\rm e}^{i\phi_{ij}(t)}c^\dag_{i\ell\sigma}\hat{T}^\alpha_{\ell\ell'}c_{j\ell'\sigma},
\label{kcuf3}
\end{align}
where $J_H$ is the Hund's coupling and $U'=U-2J_H$. The hopping matrices $\hat{T}^\alpha$'s are imposed by the cubic lattice symmetry, with $\alpha=x,y,z$ determined by the direction of the bond $\langle ij\rangle$. We solve the system again on a Bethe lattice with three types of bonds to mimic the cubic symmetry \cite{li2018nat}.

\begin{figure}
\includegraphics[scale=0.7]{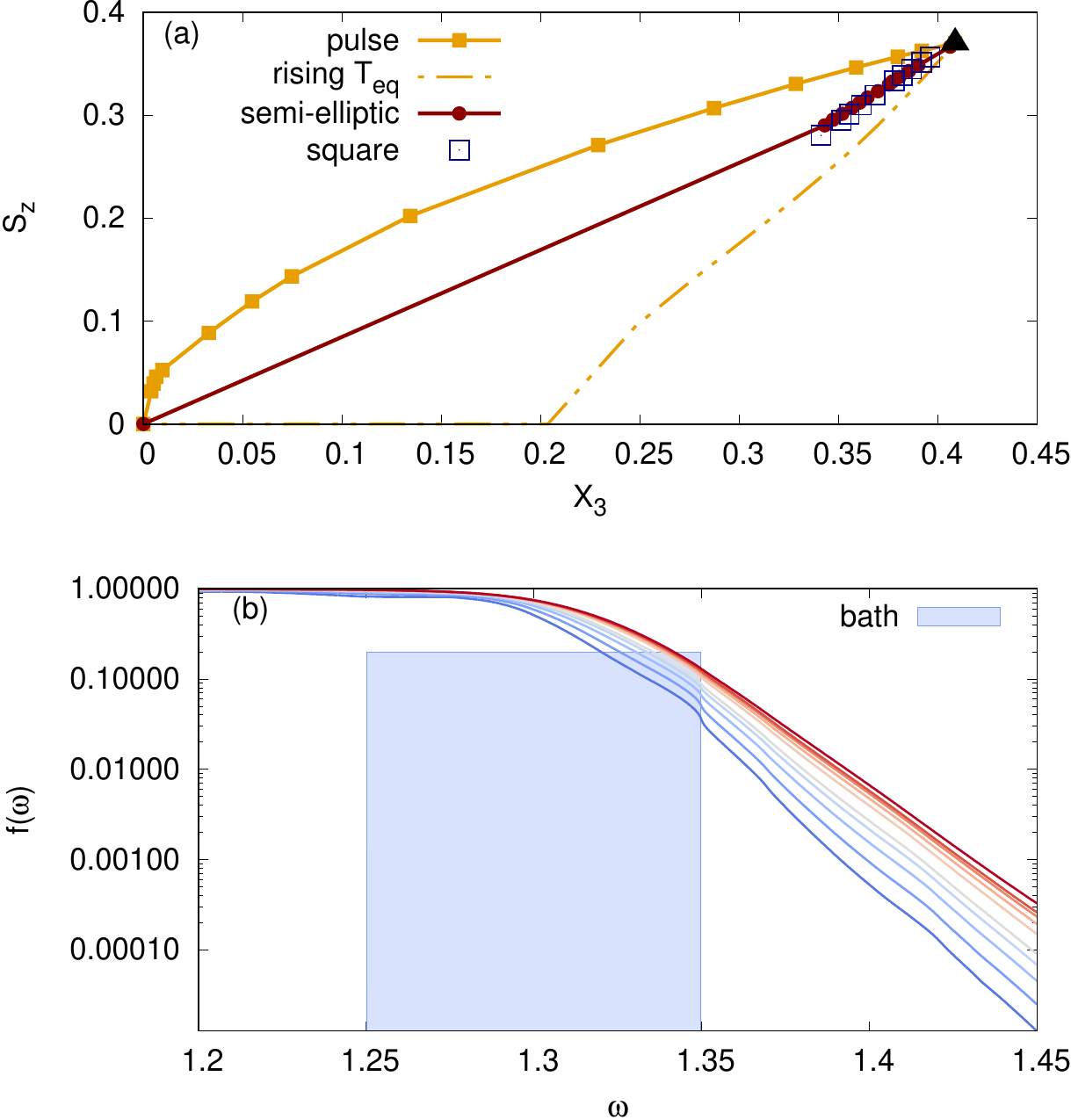}
\caption{
Intertwined spin and orbital order in a photo-doped two-band Hubbard model. (a) A-type AFM order parameter $S_z$ versus orbital order parameter $X_3$. The yellow solid line shows orders of photo-doped systems, i.e.,  after excitation with an electric field pulse. The yellow dashed line shows equilibrium order for different temperatures ($T$ increases from right to left along the curve). The red line and dots show orders in the presence of semi-elliptic baths, with $\Gamma=0.3,0.4,\ldots,1.5\times10^{-5}$. A bath bandwidth $W=0.3$ and asymmetric $V_U=1.2,V_L=-1.27$ are chosen to maintain the filling $n=3$. The blue hollow squares correspond to system coupled to square bath, with a variety of different $\Gamma$'s. $W=0.05$ and $V_U=1.3,V_L=-1.4$. The point corresponding to $\Gamma=0$ (equilibrium) is marked as a black triangle. (b) Distribution functions obtained for a coupling to square baths. The curves from blue to red, with increasing excitation density, correspond to the hollow squares in panel (a).}
\label{fig7}
\end{figure}

The ground state of \eqref{kcuf3} features an intertwined A-type AFM spin order (FM planes align antiferromagnetically), with order parameter $S_z$, and G-type antiferro-orbital order (alternating orbital occupation in all directions), with order parameter $X_3=n_1-n_2$; here $n_{1,2}$ are occupations of orbital $1$ and $2$, respectively. $X_3$ indicates the staggered polarization in orbital occupation \cite{pavarini2008}. Under non-equilibrium excitations, real-time DMFT simulations indicate that the system can evolve into a hidden phase with the ratio $S_z/X_3$ distinct from any equilibrium states \cite{li2018nat}. For comparison, we pick up the parameters from Ref.~\citenum{li2018nat}, where $U/t_0=7,J_H/U=0.1$ and equilibrium $\beta=100$. Specifically, a single-cycle electric pulse (with period $T\sim1.0/t_0$) is applied to induce a partial melting of the spin-orbital order and a non-thermal ordered state forms within about 100 hopping times (roughly 100 fs if $t_0\sim 1$ eV). By fitting the time-dependence of the two order parameters with exponential functions we obtain the extrapolated orders shown as yellow solid curve in Fig.~\ref{fig7}. One can see that the non-thermal state contains orders distinct from the configurations reached in equilibrium as a function of temperature (dotted yellow line). The non-thermal states feature \emph{stronger} A-AFM spin order than the orbital order, while the opposite situation is observed in equilibrium due to a weaker spin exchange interaction than the orbital part. This opposite behaviour comes from the interplay between charge excitations and the spin-orbital order. Again, hopping of charge excitations transfers kinetic energy to the spin and orbital orders, but in contrast to the pure antiferromagnet, hopping within the \emph{ferromagnetic} planes does not change the spin order but create trails of defects in the G-type orbital ordering, leading to faster decay of the orbital order than the spin order. 
 
While the real-time simulations confirm the non-thermal orders on several $100$ hopping times, it is not clear whether these photodoped hidden phases are intrinsically transient or can prevail until the recombination of charge excitations. In the following, we study the system using the steady-state formulation and couple it to two semi-elliptic fermion baths as in the one-band case. For three-quarter-filling, the orbital- and spin-averaged spectrum is no longer symmetric w.r.t zero frequency and the positions of the two baths need to be separately adjusted to preserve the local occupation $n=n_1+n_2=3$. We show resulting order parameters of the case $V_U=1.2,V_L=-1.27$ in Fig.~\ref{fig7}. Interestingly, under increasing bath-doping, the combined spin-orbital order becomes different from both equilibrium and the extrapolated order in the photodoped state. In fact, the $S_z/X_3$ is typically larger than the equilibrium values and smaller than the values in the transient hidden states, as seen from the red curve in the figure. We also confirm that the shape of the bath spectral density does not change the qualitative behaviour. In fact, the scenario can be best demonstrated with the case of square baths, as indicated in Fig.~\ref{fig7}(b). The decaying tails of distribution functions can be well described by exponential functions insensitive to bath details, suggesting a universal behavior as before. In addition, for different bath types (square and semi-elliptic), the two-dimensional order parameters $(S_z,X_3)$ always follow the same curve as shown in the figure.

The above result can be interpreted as the indication of a new timescale between the transient dynamics within up to hundreds of electron hoppings and the charge recombination which takes an exponentially longer time. Within this time window, the spin order can exchange energy with the orbital order, resulting in a slow relaxation between the reservoirs of the spin and orbital moments and, therefore, a spin-orbital order closer to the equilirium system. This putative new relaxational dynamics may be related to the thermalization of the Kugel-Khomskii compass model \cite{kugel1973}. 

It is worth noting that, due to the intrinsic frustration and canting of orbital ordering \cite{feiner1997}, the ultrafast dynamics in the two-band Hubbard model is typically accompanied by a precession dynamics of spin and orbital pseudospins \cite{li2018nat}. When this is considered more complicated ordering may emerge in the long time limit, such as a nonzero orbital order $X_2$, which corresponds to a complex superposition of $e_g$ orbitals that is generally absent in equilibrium \cite{maezono2000,brink2001,eckstein2017}. In the present formulation, the stationary photodoped states may be argued to contain order ``closest" to equilibrium states through adiabatically increasing $\Gamma$ from $0$. To detect unconventional orders reachable from precession dynamics in the photoexcited systems, one could try to scan the phase diagram by putting a proper seed for respective symmetry breaking, or by computing the relevant susceptibilities {\em of the nonequilibrium steady state} towards other instabilities.

\section{Conclusion}
\label{conclusion}
In this article, we discussed an alternative numerical method to compute the quasi-stationary behaviors of the photodoped Mott insulators. Specifically, since the timescale of charge recombination in Mott insulators is well beyond those of other relevant processes, a stationary photodoped state can be stabilized with a very weak external driving, such as a weak coupling to certain fermion baths, while producing quantitatively identical physical properties to its transient counterpart. Using DMFT combined with a steady-state NCA solver, we demonstrated that the bath-coupling protocol is well-defined, so that the resulting stationary photodoped states are generally insensitive to the bath details. We systematically compared their physical properties with the transient photodoped states created by laser excitations. In particular, we find that the manifold of stationary photodoped states can be parametrized by a few physical quantities, i.e., the effective temperature $T_{\rm eff}$ and the density of charge excitations $n_{\rm ex}$. Indeed, with the same parameters, a stationary photodoped state exhibits the same spectral features and charge distribution as a transient (quasi-stationary) photodoped state. 

Using this steady-state theory, we examined the long-time behavior of a photodoped two-band Mott insulator, relevant to transition metal compounds like KCuF\textsubscript{3}. Intriguingly, we find a new nonthermal spin-orbital order under photodoping, indicated by a ratio $S_z/X_3$ which is distinct from either equilibrium states or transient states obtained from short-time DMFT simulations. This hidden state may be attributed to the equilibration between spin and orbital moments, possibly through superexchange mechanism, corresponding to a timescale between the short-time dynamics (up to hundreds of fs) and the possible eventual thermalization towards an equilibrium state, dominated by the charge recombination.

The steady-state theory holds the promise of resolving non-thermal effects and hidden phases of multi-band Mott insulators in the long-time limit with reduced computational efforts. It provides a powerful tool for future studies in many directions, such as photoinduced superconductivity \cite{li2019, werner2019}, the entangled dynamics between lattice and charge \cite{grandi2020}, and strongly correlated systems driven by quantum light \cite{mazza2019,sentef2020,li2020}. A $GW$+EDMFT scheme with the steady-state setup can also be promising to study the charge transfer dynamics in photodoped states \cite{golez2015,golez2018,golez2019}. The slow dynamics of the quasi-steady states could be addressed by a suitable quantum kinetic theory for correlated systems  \cite{wais2018}. 

\begin{acknowledgments}
We thank Y.~Murakami, P.~Werner, M.~Sentef, D.~Gole\ifmmode \check{z}\else \v{z}\fi{}, and H.~Strand for helpful discussions. We were supported by ERC Starting Grant No. 716648.
\end{acknowledgments}
\appendix
\section{The steady-state implementation of NCA}
In this section, we provide details of the implementation of the steady-state NCA. In a steady-state, all physical observables are stationary in time and the Green's functions are time translational invariant $G(t,t')=G(t-t')$. Thus, we consider Green's functions defined on the Keldysh contour $\mathcal{C}$ where the initial correlation is decoupled from the relevant time evolution \cite{aoki2014}. 

To clarify the non-crossing approximation used in this article, we briefly summarize the strong-coupling expansion on the Keldysh contour\cite{eckstein2010} in the following. We embed the impurity problem \eqref{imp} into a larger Hilbert space of pseudo-particles where each pseudo-particle $f_p$ is mapped from a unique local basis state $|p\rangle$ in the original problem. For the one-orbital case in particular, we have $p\in\{0,\uparrow,\downarrow,\uparrow\downarrow\}$, where $0$ represents the vacuum state. We can, therefore, define the pseudo-particle Green's functions and impose the physical constraint $Q=\sum_p f^\dag_pf_p=1$ by projecting all physical quantities into the $Q=1$ subspace. The \emph{projected} pseudo-particle Green's functions satisfy the following equation of motion on the Keldysh contour,
\begin{align}
[i\partial_t-H_{\rm loc}]\mathcal{G}(t,t')-\int_{\mathcal{C},t'\prec \bar{t}\prec t}d\bar{t}\Sigma(t,\bar{t})\mathcal{G}(\bar{t},t')=\delta_{\mathcal{C}}(t,t'),
\label{eom_nca}
\end{align}
where $\prec$ denotes cyclic order on the contour and $H_{\rm loc}$ is the local Hamiltonian expanded in the local basis $|p\rangle$. The self-energy $\Sigma(t,t')$ comes from the hybridization of the local impurity and the self-consistent bath. 

\subsection{Formulation for the steady-state}
We first rewrite the equation of motion in a form that is suitable for the steady-state problem, in which the Green's functions only rely on the relative time $t-t'$. The key observation is that the cyclic convolution in Eq.~\eqref{eom_nca} can be simplified by defining retarded and advanced components $\mathcal{G}^r(t,t')=\theta(t-t')\mathcal{G}^>(t,t')$ for $t>t'$ and $\mathcal{G}^a(t,t')=-\theta(t'-t)\mathcal{G}^<(t,t')$ for $t<t'$, where $G^<$ and $G^>$ are usual lesser and greater components \cite{murakami2018}. It is straightforward to verify that these Green's functions satisfy the usual hermiticity condition and the analogous Langreth rules for cyclic convolution. In fact, we find that
\begin{align}
[\mathcal{G}_1*\mathcal{G}_2]^r(t,t')&=\int_{t'}^{t}d\bar{t}\mathcal{G}^r_1(t,\bar{t})\mathcal{G}^r_2(\bar{t},t')\nonumber\\
&=\int_{-\infty}^{\infty}d\bar{t}\mathcal{G}^r_1(t,\bar{t})\mathcal{G}^r_2(\bar{t},t')\\
[\mathcal{G}_1*\mathcal{G}_2]^<(t,t')&=-\int_{t'}^{-\infty}d\bar{t}\mathcal{G}^<_1(t,\bar{t})\mathcal{G}^a_2(\bar{t},t')\nonumber\\
&+\int_{-\infty}^{t}d\bar{t}\mathcal{G}^r_1(t,\bar{t})\mathcal{G}^<_2(\bar{t},t')\nonumber\\
&=\int_{-\infty}^{\infty}d\bar{t}[\mathcal{G}^r_1(t,\bar{t})\mathcal{G}^<_2(\bar{t},t')\nonumber\\
&+\mathcal{G}^<_1(t,\bar{t})\mathcal{G}^a_2(\bar{t},t')],
\end{align}
where $*$ is the cyclic convoluation. From this observation, one immediately obtains the following form of the equation of motion \eqref{eom_nca},
\begin{align}
&[i\partial_t-H_{\rm loc}]\mathcal{G}^{r}(t,t')-\int_{-\infty}^{\infty}\Sigma^r(t,\bar{t})\mathcal{G}^{r}(\bar{t},t')=0,\nonumber\\
&\mathcal{G}^<(t,t')=\int_{-\infty}^{\infty}dsds'\mathcal{G}^{r}(t,s)\Sigma^<(s,s')\mathcal{G}^{a}(s,t').
\label{ggwe33p3l3}
\end{align}
These equations can be transformed to the frequency domain and computed efficiently. Within the non-crossing approximation, i.e., the lowest order self-consistent approximation for the action \eqref{imp}, the self-energies are given by
\begin{align}
\Sigma_{0}(t)&=-i\sum_\sigma\Delta_\sigma(-t)\mathcal{G}_\sigma(t),\nonumber\\
\Sigma_\sigma(t)&=i\Delta_\sigma(t)\mathcal{G}_0(t)-i\Delta_{\bar{\sigma}}(t)\mathcal{G}_{\uparrow\downarrow}(t),\nonumber\\
\Sigma_{\uparrow\downarrow}(t)&=i\sum_\sigma\Delta_\sigma(t)\mathcal{G}_{\bar{\sigma}}(t).
\label{nsm333h3j3}
\end{align}
Here we have assumed a spin-diagonal hybridization function for simplicity. A generalization of the digarmmatic expressions to more orbitals, with spin  and orbitally off-diagonal  hybridization functions is straightforward.\cite{eckstein2010}. The impurity Green's functions can be calculated by similar diagrammatic expressions,
\begin{align}
\label{ehjdkecld;w}
G_{\sigma}(t)=i[\mathcal{G}_\sigma(t)\mathcal{G}_0(-t)-\mathcal{G}_{\uparrow\downarrow}(t)\mathcal{G}_{\bar{\sigma}}(-t)]/Q,
\end{align}
with normalization factor $Q=\sum_p(-1)^p\mathcal{G}_{pp}^<(0)$. In practice, one usually adds a pseudoparticle chemical potential $\lambda$, which is determined self-consistently during the iterations, in the local hamiltonian $H_{\rm loc}$ to normalize $Q=1$ and help convergence. Because NCA is a diagrammatic expression in terms of the full propagators, the pseudo-particle Dyson equations \eqref{ggwe33p3l3} and the diagrammatic equations \eqref{nsm333h3j3} and \eqref{ehjdkecld;w} have to be solved self-consistently. In the steady state code, this is achieved by iteration (see below).

\subsection{Taming the numerical instability at low temperature}
At low temperatures, one has to use a very fine frequency grid to stabilize the calcuation. A logarithmic grid can be used to get around this problem, but it then loses the advantage of the fast Fourier transform. We therefore use an equally spaced grid in this paper. Another subtlety comes from the fact that, in the first iteration of the self-consistent equastions \eqref{ggwe33p3l3} to \eqref{ehjdkecld;w}, the self energies are unknown and we often start with the Green's functions of an isolated impurity,
\begin{align}
\mathcal{G}^r(\omega)&=(\omega - H_{\rm loc} - i0^+)^{-1},
\label{freq_gret}\\
\mathcal{G}^<(\omega)&=\zeta \left(\mathcal{G}^r(\omega)-\mathcal{G}^a(\omega)\right){\rm e}^{-\beta\omega},
\label{freq_gles}
\end{align}
where $\zeta$ is a diagonal matrix with element $\pm1$ corresponding to boson and fermion pseudo-particles, respectively \cite{eckstein2010}. The ``fluctuation-dissipation theorem" for psudo-particle propagators differs from that of regular Green's functions, and features an infrared divergence of the factor ${\rm e}^{-\beta\omega}$ as $\omega \to-\infty$, which  should be treated with extra care. Although this is superficial in theory due to the infrared threshold behaviour of pseudo-particle spectrum \cite{coleman1984}, it can nevertheless lead to numerical instability at low temperatures. This problem can be avoided by replacing $0^+$ by $\eta[1- f(\beta\omega)]$ with $f(x)=1/({\rm e}^x+1)$ with $\eta\to0$, which normalizes the exponential factor to a well-behaved factor $f(\beta\omega)$ in Eq.~\eqref{freq_gles}, while preserving the same fluctuation-dissipation theorem. At very low temperatures, the pseudoparticle spectrum can contain sharp peaks near the chemical potential $\lambda$, and the $\eta$ factor may be added as a regulator in the Dyson's equation for all iterations.

\section{The dependence on initial guesses}
\begin{figure}
\includegraphics[scale=0.7]{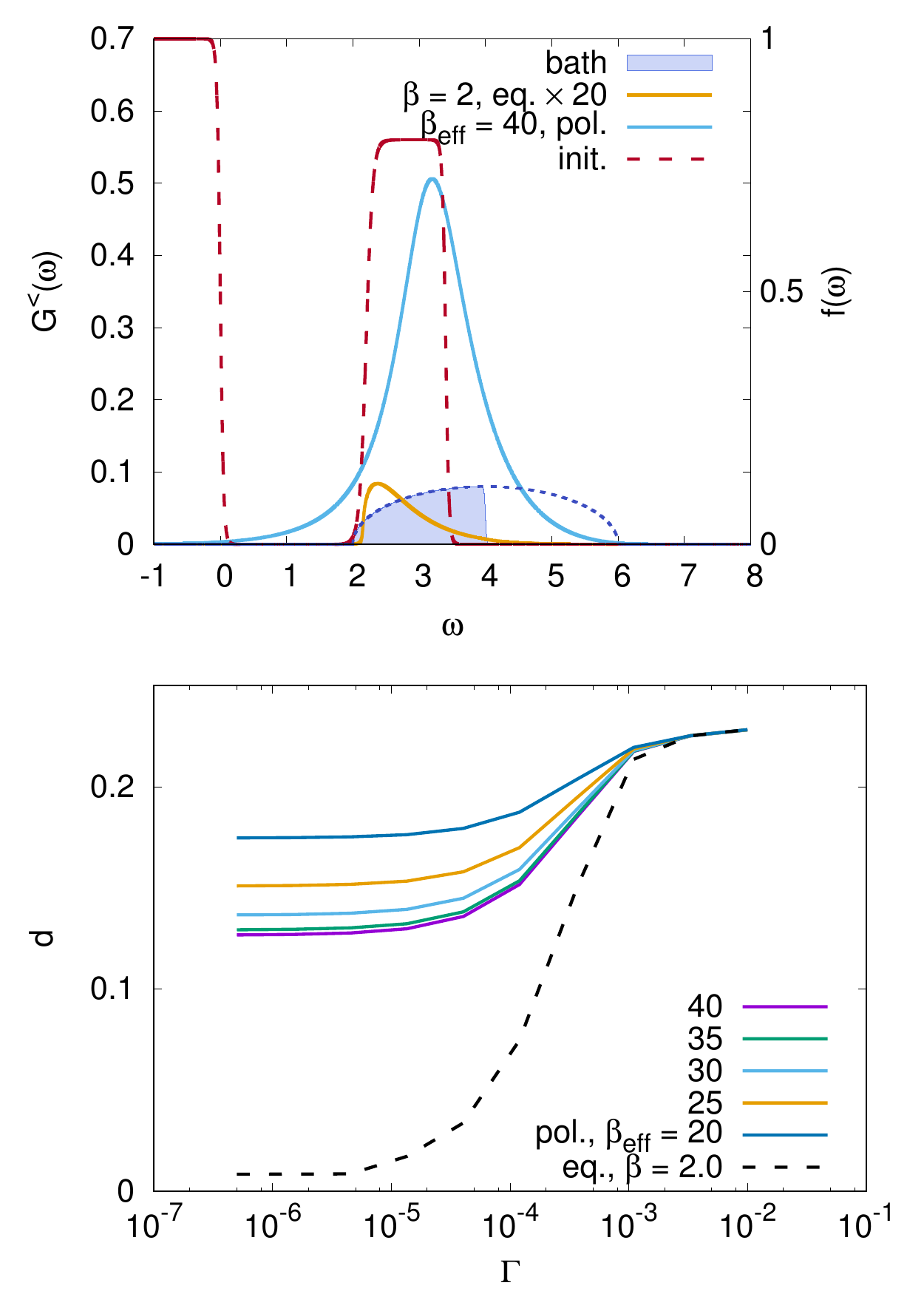}
\caption{Non-equilibrium steady states pumped up by the fermion baths. In the left panel, the dashed line indicates an equilibrium initial guess, while the solid lines correspond to a ``polarized" initial guess. In the right panel, the red dashed line shows the distribution function of the polarized initial state. The blue dashed line shows the fermion-bath spectrum (in the upper Hubbard band) and the blue shaded area shows its occupation.}
\label{figA1}
\end{figure}

In the main text, we have concentrated on relatively small photodoping (up to several percents for double occupancy), which is physically close to equilibrium states. In this appendix, we show that, when states become extremely (if not unphysically) far from equilibrium, and when the bath-coupling is extremely small, the solution can depend sensitively on the initial guess. We consider a paramagnetic phase of the one-band Hubbard model of $U=8t_0$ coupled to fermion baths, as shown in Fig.~\ref{figA1}. Exemplarily, the calculations are started with two initial guesses, the equilibrium ground state with a finite inverse temperature $\beta$, and a ``polarized" state with a photodoped-like distribution function as shown in the right panel. Specifically, in the polarized case, a peak is added to the Fermi-Dirac function of inverse temperature $\beta_{\rm eff}$ at about $\omega\sim2.5$, and a dip antisymmetric to the peak is added at about $\omega=-2.5$ (not shown in the figure), with the half-filling condition $f_{\rm eff}(-\omega) + f_{\rm eff}(\omega) = 1.0$ preserved. The upper edge of the peak at $\omega\sim 3.5$ is of the Fermi-Dirac form with the same $\beta_{\rm eff}$. 

For the two cases, we check the double occupancy $d=\langle n_{\uparrow}n_{\downarrow}\rangle$, which reflects the photodoping level in the system. From the view of pumping up approximately integrable systems \cite{lange2017}, the approximately conserved $d$ is pumped by the bath-coupling, and can be in principle driven to a very large value even with negligible $\Gamma$. Here, the upper fermion bath is chosen to be roughly half-filled, with a chemical potential $\mu_U\sim4.0$, and the lower fermion bath is symmetric to it. As expected, $d$ increases with bath coupling $\Gamma$. At $\Gamma\gtrsim 10^{-3}$, the two initial guesses lead to almost identical double occupancy. It is intriguing that, in the small $\Gamma$ limit, the solutions from different initial guesses start to deviate, indicating the coexistence of different photodoped states for the same parameters. With the polarized initial guess, the double occupancy approaches a large value $d>0.1$ in the limit $\Gamma\to0$. The value further depends on the effective temperature of the initial guess.

Note that, without the fermion bath coupling, both initial guesses lead to completely thermal solutions after convergence. Here, the results show that an extremely nonequilbrium state can be maintained by a negligible bath-coupling, and the converged solution in our calculations is not necessarily unique. Physically, it is of course expected that two physical phases can coexist for the same parameter set, such as in a first-order dynamical phase transition of the NESS \cite{li2015}. Furthermore, multiple physical solutions can stay close to each other in the manifold of photodoped states, depending sensitively on numerical precision and the initial guess. It is, therefore, important to check whether these states indeed approximate some physical transient states for our purpose.
\bibliography{ness.bib}
\end{document}